\documentclass[%
 reprint,
 amsmath,amssymb,
 aps,
]{revtex4-2}

\usepackage{graphicx}
\usepackage{dcolumn}
\usepackage{bm}

\usepackage{color}
\usepackage{physics}

\def\cm{cm$^{-1}$}

\begin{document}

\title{Exfoliation and Optical Properties of S=1 Triangular Lattice Antiferromagnet NiGa$_2$S$_4$ }

\author{Jazzmin Victorin$^1$, Aleksandar Razpopov$^2$, Tomoya Higo$^{3,4}$, Reynolds Dziobek-Garrett$^5$, Thomas J. Kempa$^5$, Satoru Nakatsuji$^{3,4,1,6}$, Roser Valent\'\i$^2$, Natalia Drichko$^{1,*}$}

\affiliation{$^1$Institute for Quantum Matter and Department of Physics and Astronomy, Johns Hopkins University, Baltimore, Maryland 21218, USA}

\affiliation{$^2$Institute for Theoretical Physics, Goethe University Frankfurt, Max-von-Laue-Strasse 1, 60438 Frankfurt am Main, Germany}

\affiliation{$^3$Institute for Solid State Physics, University of Tokyo, Kashiwa, Chiba 277-8581, Japan}

\affiliation{$^4$CREST, Japan Science and Technology Agency, Kawaguchi, Saitama 332-0012, Japan}

\affiliation{$^5$Department of Chemistry, Johns Hopkins University, Baltimore, Maryland 21218, USA}

\affiliation{$^6$Canadian Institute for Advanced Research (CIFAR), Toronto, Ontario M5G 1M1, Canada}

\affiliation{$^{*}$ Corresponding author's email: drichko@jhu.edu}

\begin{abstract}

Two-dimensional (2D) van der Waals (vdW) materials have been an exciting area of research ever since scientists first isolated a single layer of graphene. Single layer magnetic materials can provide a pathway for vdW heterostructures with magnetic properties. While most of the magnetic vdW materials exhibit ordering transitions in the bulk, here we report a successful exfoliation of a triangular lattice S=1 antiferromagnet    NiGa$_2$S$_4$, which already demonstrates exotic magnetism in the bulk material. We establish the number of layers of the material by atomic force microscopy (AFM) and detail a careful characterization using Raman and optical spectroscopy to demonstrate how the optical, electronic, and structural properties of NiGa$_2$S$_4$ change as a function of sample thickness. Optical measurements and electronic structure calculations of bulk versus monolayer NiGa$_2$S$_4$ confirm the material to be a Mott insulator with an electronic gap of about 1.5 eV, which slightly increases for layers below 10 L. We conclude with a theoretical analysis of the possibility of doping monolayer NiGa$_2$S$_4$ by proximity to a metal.
    
\end{abstract}


\maketitle
 
\section{Introduction}

Magnetic two-dimensional (2D) van der Waals (vdW) materials have quickly become a promising area of condensed matter research for their multitude of exciting properties and potential applications ranging from quantum computing, topological magnonics, low-power spintronics, and optical communication\cite{2DMagnets}. However, despite large interest the number of realizations of magnetic 2D materials is rather limited, with most of the materials characterized so far showing a transition to magnetically ordered states at relatively high temperatures. In contrast, the magnetism in bulk S=1 triangular lattice antiferromagnet NiGa$_2$S$_4$ has already attracted attention due to exotic magnetic properties: no magnetic ordering is detected despite magnetic interactions on the order of meV, but unconventional spin freezing is observed around 8~K.~\cite{spin_nakatsuji_2005,Valentine2020,Nambu2015,Stock2010}. This behavior was discussed in terms of frustrated interactions and bi-quadratic coupling in S=1 system~\cite{Stoudenmire2009,Takano2011,Bhattacharjee2006,Lauchli2006}. Exfoliation of such material is of an applied interest, and can provide a pathway to answering questions related to exotic magnetism in  NiGa$_2$S$_4$.

In this work we demonstrate that it is possible to produce at least two-layer thick flakes of S = 1 triangular lattice antiferromagnet NiGa$_2$S$_4$ \textit{via} mechanical exfoliation. We present a characterization of the exfoliated flakes with atomic force spectroscopy (AFM), Raman scattering spectroscopy, and optical spectroscopy. We compare the experimentally measured optical gap to the calculations of  the electronic properties of bulk and single layer NiGa$_2$S$_4$ in the framework of density functional theory (DFT). Anticipating the interest in this vdW magnet as a part of vdW heterostructure assembly, we discuss the possibility of doping of a single layer of NiGa$_2$S$_4$ by proximity to a substrate. A highly frustrated vdW material can open new possibilities for properties of vdW heterostructures including the possibility of injecting charge carriers into a single layer of a S = 1 triangular lattice vdW system with strong magnetic correlations. In fact, the first step to test an application of 15~$\mu$m flakes of NiGa$_2$S$_4$  as photovoltaic detectors demonstrated changes of optical properties related to doping the material by electrolyte solution~\cite{Serra2023}. 

Exfoliation and studies of thin flakes of NiGa$_2$S$_4$ can also answer questions related to magnetism in the bulk. Local broken translational symmetry brought about by stacking faults along the $c$-axis, known to exist in the bulk material~\cite{stackingfaults}, has been cited as a potential cause for the disorder preventing magnetic ordering. We show that by producing flakes of NiGa$_2$S$_4$ below 50 layers in thickness, we eliminate stacking faults, resulting in a cleaner system wherein this systems' magnetic properties can be studied.     

\section{Results}\label{sec2}

\begin{figure}
    \includegraphics[width=\linewidth]{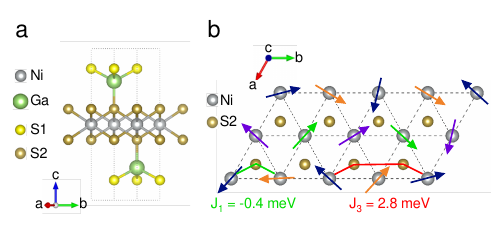}
    \caption{\textbf{Crystal structure of NiGa$_2$S$_4$.} (a) Dashed lines indicate the unit cell of NiGa$_2$S$_4$ and a single layer of the material. (b) Schematic view of the triangular lattice of Ni$^{2+}$ with J$_1$ and J$_3$ magnetic superexchange interactions through S2 atoms (inside sulphurs) depicted and schematic view of short-range order ~\cite{Valentine2020}}
    \label{Fig.crystal structure}
\end{figure}

NiGa$_2$S$_4$ has a layered crystal structure of edge-sharing NiS$_6$ octahedra forming a 2D isotropic triangular lattice of S=1 Ni$^{2+}$ in the $ab$ plane. Layers of GaS$_4$ tetrahedra are positioned on either side of the Ni layer along the c axis. In Fig.~\ref{Fig.crystal structure}, S1 refers to the ``outside" sulfurs bound to the Ga atoms while S2 refers to the ``inner" sulfurs of the Ni octahedral environment. The crystal has a trigonal symmetry with a = b = 3.62680~\AA~, c = 12.00180~\AA~, $\alpha$ = $\beta$ = 90$^\circ$, and $\gamma$ = 120$^\circ$~\cite{stackingfaults}, belonging to the P$\bar{3}$m1 space group with D$_{3d}$ point group symmetry. The layers have only vdW interactions along the $c$-axis, which allows for mechanical exfoliation of the material. This also results in the stacking faults detected in the bulk which break translational symmetry in the material~\cite{stackingfaults}. 

\subsection{Preparation and characterization of atomically-thin layers using Atomic Force Microscopy}\label{AFM}

\begin{figure}
    \includegraphics[width=\linewidth]{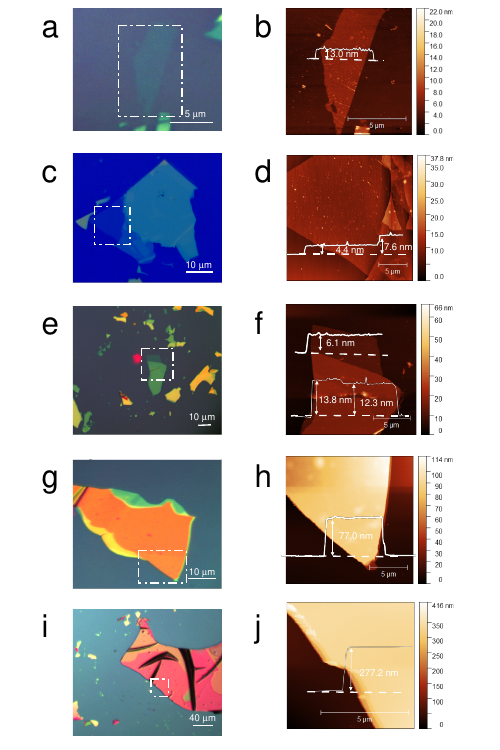}
    \caption{\textbf{Optical images and AFM characterization of mechanically exfoliated NiGa$_2$S$_4$ flakes} Left column shows images obtained with an optical microscope, right column presents AFM images with height profile. (a)  Dashed white box area indicates bilayer area of flake. (b) AFM of enclosed area in (a). Inset: height profile of flake. (c)-(d) Optical and AFM image of 3L and 5L area of flake. (e)-(f) Optical and AFM image of 4L, 8L, and 9L areas of an exfoliated flake. (g)-(h) Optical and AFM image of 50L flake. (i)-(j) Optical and AFM image of 180L flake.}
    \label{Fig.AFM}
\end{figure}

 \begin{figure*}
    \includegraphics[width=\linewidth]{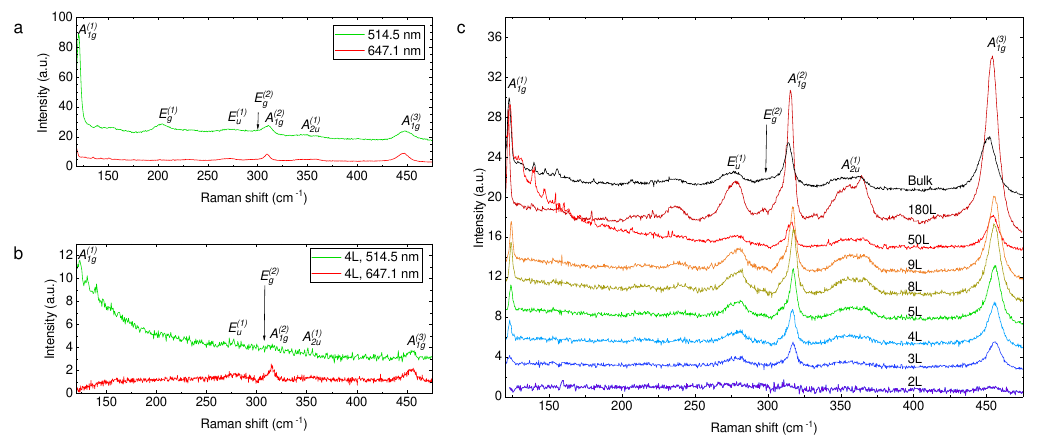}
    \caption{\textbf{Raman spectra of exfoliated NiGa$_2$S$_4$} (a) Non-polarized Raman spectra, (x,x) + (x,y), of bulk NiGa$_2$S$_4$ using $\lambda_{exc}$ = 514.5 nm and 647.1 nm. (b) Non-polarized Raman spectra, (x,x) + (x,y), of 4L NiGa$_2$S$_4$ flake using $\lambda_{exc}$ = 514.5 nm and 647.1 nm. (c) Non-polarized Raman spectra, (x,x) + (x,y), of bulk NiGa$_2$S$_4$ and exfoliated flakes vertically offset for clarity using $\lambda_{exc}$ = 647.1 nm.}
    \label{Fig.Raman}
\end{figure*}

 Few-layer NiGa$_2$S$_4$ flakes were mechanically exfoliated from the bulk crystal onto silicon substrates capped with 300 nm of silicon dioxide (SiO$_2$) using the traditional mechanical exfoliation method. In Fig.~\ref{Fig.AFM}, we present optical microscope images (left column) and atomic force microscope (AFM) images (right column) of the exfoliated flakes.   

 Fig.~\ref{Fig.AFM}(a) and (b) show images of the thinnest flake characterized. The dashed white box area in the optical microscope image in Fig.~\ref{Fig.AFM}(a) indicates the area of interest. The topography of the flake obtained by AFM marked by the dashed white box area is shown in Fig.~\ref{Fig.AFM}(b) with the height profile of the flake inset on the image. The height obtained by AFM indicates the thinnest flake characterized in this study is 3 nm thick. Given the crystal structure of NiGa$_2$S$_4$, the estimated thickness of a monolayer is approximately 1.4 nm. This value is calculated as a sum of the $c$ lattice parameter, 1.2 nm, and a typical value of 0.2 nm to account for the distance between the flakes and the Si/SiO$_2$ wafer. Such an offset is typically attributed to the presence of adsorbates\cite{Substrateoffset} trapped between the exfoliated flakes and the substrate as well as the force the AFM tip applies to the exfoliated flakes. Using this expectation value for a monolayer, the flake depicted in Fig.~\ref{Fig.AFM}(b) was determined to be two layers thick, taking into account the interlayer distance. The difference in height between an 8L and 9L flake, seen in Fig.~\ref{Fig.AFM}(f), indicates this interlayer distance is about 0.31 nm.  
 
 In Fig.~\ref{Fig.AFM}(c) and \ref{Fig.AFM}(d), an optical image of a 3L and 5L flake is shown in the dashed white box along with its accompanying AFM image. The inset in \ref{Fig.AFM}(d) shows the height profiles for both the three layer area, which has a measured height of 4.42 nm, and five layer area, which has a measured height of 7.57 nm. In Fig.~\ref{Fig.AFM}(e) and \ref{Fig.AFM}(f), an optical image of a 4L flake is shown in the dashed white box along with its accompanying AFM image depicting a measured height of 6.07 nm. Fig.~\ref{Fig.AFM}(g) and \ref{Fig.AFM}(h) show the images of a 50L flake, which has a measured height of 77 nm. Note the orange color of the flake. Reflectivity measurements and calculations of optical constants presented below show that the color change is a result of the flake thickness becoming lower than penetration depth of light, and an interference in the layers of Si/SiO$_2$ and the flake. Fig.~\ref{Fig.AFM}(i) and \ref{Fig.AFM}(j) show the images of a flake with about 180 layers, which has a measured height of 277.2 nm. 

\subsection{Microscopic characterization using Raman Spectroscopy}\label{Raman}

 \begin{figure*}
    \includegraphics[width=\linewidth]{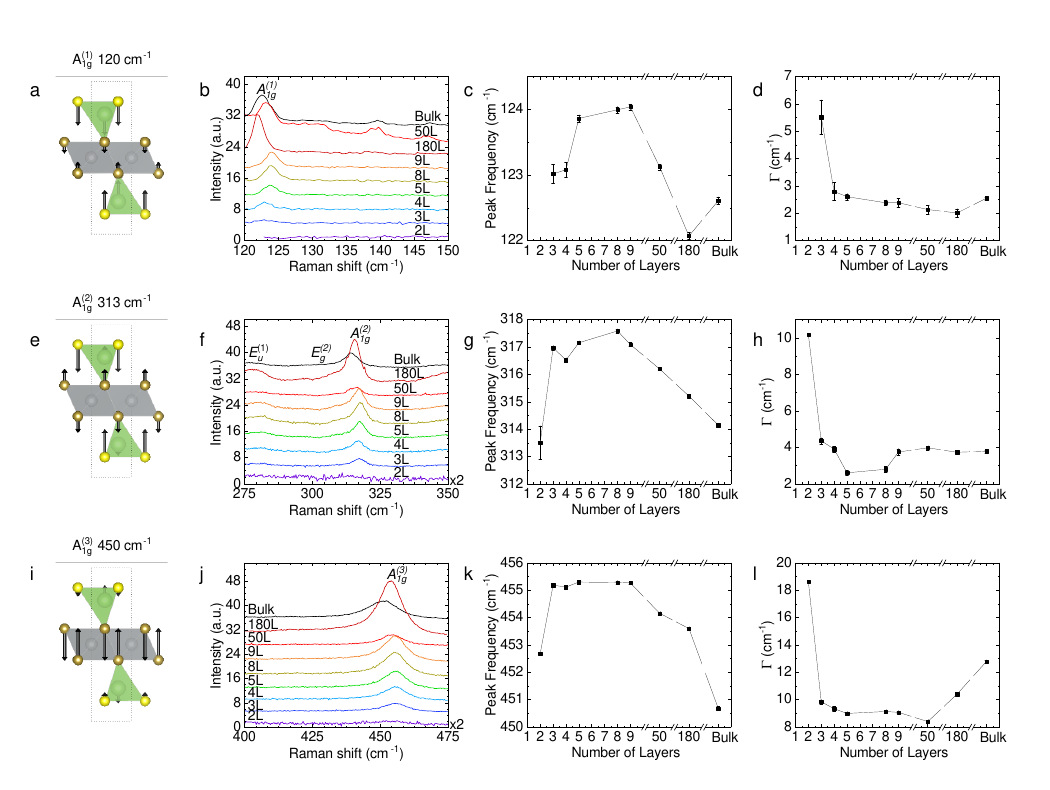}
    \caption{\textbf{Raman spectra of out-of-plane A$_{1g}$ phonons for exfoliated flakes of NiGa$_2$S$_4$ }. Left column presents schematic of vibration of phonons, second column shows spectra of the flakes in the relevant range, third column shows phonon frequencies as a function of thickness, and the last column shows width of phonon modes as a function of thickness. (a-d) A$^{(1)}_{1g}$ 120 cm$^{-1}$ phonon. (e-h) Same figures for A$^{(2)}_{1g}$ 313 cm$^{-1}$ phonon. (i-l) Same figures for A$^{(3)}_{1g}$ 450 cm$^{-1}$ phonon.}
    \label{Fig.A1g phonons}
\end{figure*}

\begin{figure*}
    \includegraphics[width=\linewidth]{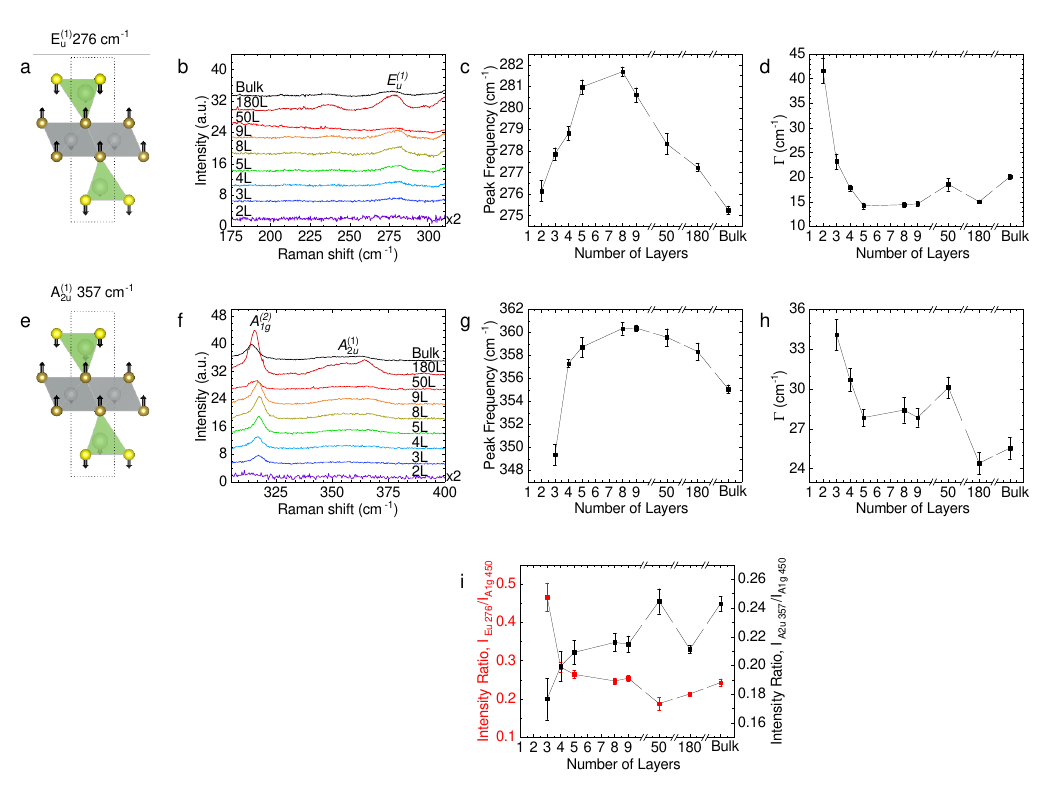}
    \caption{\textbf{Behavior of forbidden IR modes} (a) Schematic of vibration of E$^{(1)}_{u}$ 276 cm$^{-1}$ phonon. (b) Raman spectra of bulk and few-layer NiGa$_2$S$_4$ centered around E$^{(1)}_{u}$ 276 cm$^{-1}$ vertically offset for clarity. (c) Peak center frequency of E$^{(1)}_{u}$ 276 cm$^{-1}$ phonon as a function of sample thickness. (d) Linewidth of E$^{(1)}_{u}$ 276 cm$^{-1}$ phonon as a function of sample thickness. (e-h) Same figures for A$^{(1)}_{2u}$ 357 cm$^{-1}$ phonon. (i) Ratio of E$^{(1)}_{u}$ intensity to A$_{1g}$ 450 cm$^{-1}$ intensity and ratio of A$^{(1)}_{2u}$ intensity to A$_{1g}$ intensity as a function of sample thickness.}
    \label{IRActiveModes}
\end{figure*}

Micro-Raman spectroscopy has become a standard tool for characterization of vdW materials down to a single layer with the dependence of phonon frequencies on the number of layers~\cite{Lee2010,Cong2020}. In this work, we use micro-Raman spectroscopy to characterize exfoliated flakes of NiGa$_2$S$_4$ in order to establish the dependence of the number of layers on the $\Gamma$-point phonon frequencies. 

In Fig.~\ref{Fig.Raman}(a), we show spectra of bulk NiGa$_2$S$_4$ using $\lambda_{exc}$ = 514.5 nm and $\lambda_{exc}$ = 647.1 nm. Not only is the intensity of the bulk spectra higher with $\lambda_{exc}$ = 514.5 nm, but the E$_g^{(1)}$ phonon is only present in the spectra excited with  $\lambda_{exc}$ = 514.5 nm. A similar resonance enhancement of the spectra was observed in other materials containing Ni-S octahedra as building blocks~\cite{KNiPS4_1995,NiPS3}. Ref.~ \cite{NiPS3} assigns it to a coupling to electronic transitions between orbitals in the Ni environment.

The intensity of the Raman scattering in the exfoliated flakes excited with $\lambda_{exc}$ = 514.5 nm was dramatically reduced in comparison to the flakes excited with $\lambda_{exc}$ = 647.1 nm, see Fig.~\ref{Fig.Raman}(b). Therefore, an excitation wavelength of 647.1 nm was used to characterize the exfoliated flakes presented in this paper. This dramatic reduction in intensity could be initial evidence of a change in band structure from the bulk crystal to the atomically thin flakes. Since an excitation wavelength of 647.1 nm was used, the E$^{(1)}_g$ 206 cm$^{-1}$ could not be characterized due to selective resonance enhancement mediated by electron-phonon coupling\cite{NiPS3}.

The irreducible representation of zone center phonons in NiGa$_2$S$_4$, $\Gamma$ = 3A$_{1g}$ + 3E$_g$ + 4A$_{2u}$ + 4E$_u$, predicts six Raman active modes and eight infrared (IR) active modes. The assignment of the phonon frequencies, as well as findings related to a large intensity of IR-active phonons in Raman spectra were presented in Ref.~\cite{Valentine2020}. Both the A$^{(1)}_{1g}$ phonon at 120 cm$^{-1}$ and the A$^{(2)}_{1g}$ at 313 cm$^{-1}$ involve out-of-plane motion of ``outer'' sulfur atoms, S1, connected to the Ga atoms. In analogy to the layer-dependent Raman spectra of MoS$_2$, these phonons are expected to be the most sensitive to the number of layers~\cite{Lee2010}.  
The motion of ``inner'' sulfur atoms, S2, contributes largely to the A$^{(3)}_{1g}$ phonon at 450 cm$^{-1}$, see Fig.~\ref{Fig.A1g phonons} for the scheme of the phonons eigenvectors~\cite{Valentine2020}. 

In Fig.~\ref{Fig.Raman}(c), we present spectra of bulk and exfoliated NiGa$_2$S$_4$ flakes of 180, 50, 9, 8, 5, 4, 3, and 2 layers in the range of 120 cm$^{-1}$ to 475 cm$^{-1}$. The spectra are offset vertically for clarity. The schemes of the phonon eigenvectors and the dependence of the phonon parameters on the number of layers are presented in Fig.~\ref{Fig.A1g phonons}. 

While flakes of the thickness of tens of layers are expected to have spectra similar to bulk materials, we find that Raman spectra of such thick flakes differ from bulk in this material, as demonstrated by the spectrum of 180L and 50L flakes. All of the phonons associated with the out-of-plane motion of the ``outer" sulfur atoms in the 50-layer spectra shift to higher frequencies. The shift is different for different modes, excluding a systematic error. Surprisingly, the largest change is observed on the A$^{(3)}_{1g}$ phonon at about 450~\cm, which is associated with the out-of-plane motion of the ``inner" sulfur atoms. The phonon shifts to higher frequencies and narrows considerably, with the full-width-at-half-max (FWHM) $\Gamma$ going from 14 to 9 \cm. The intensity of Raman spectra decreases for thicker flakes about 50L compared to the bulk and 180L, but then increases again for 9L, with the spectra of 9L-thick flakes showing intensity higher than that of the bulk sample. Below we discuss how optical properties and electronic structure of NiGa$_2$S$_4$ determines the intensities.  

For the thin flakes, the strongest changes on the phonons are observed for the 3L and 2L flakes. All phonons show softening by few wavenumbers and broadening on exfoliation down to 3 layers, and the phonon which still have observable frequencies become essentially broader in the spectra of 2L flakes (Fig. \ref{Fig.A1g phonons}(k-l)). The most consistent changes starting with 5L flakes are observed for the A$^{(1)}_{1g}$ 120 cm$^{-1}$ phonon, which involves the movement of ``outer'' S1 and Ga atoms. The mode softens and broadens for the thin layers.

Raman spectra of the bulk crystals of NiGa$_2$S$_4$ show a non-negligible intensity of IR-active (Raman forbidden) modes, which increase in intensity upon cooling. Such Raman activity of E$^{(1)}_u$ 276 cm$^{-1}$ and A$^{(1)}_{2u}$ 357 cm$^{-1}$ IR modes were assigned to the local loss of inversion symmetry, which can have two possible origins, sulfur vacancies or stacking faults~\cite{Valentine2020}. For the latter, exfoliation could result in a cleaner system, where the intensity of the forbidden IR modes would decrease in comparison to the Raman active phonons as a function of sample thickness. In order to test this, the ratio of the forbidden IR active modes intensity to the Raman active modes intensity was taken and plotted as a function of sample thickness in Fig.~\ref{IRActiveModes}(i). Interestingly, we see that while the A$^{(1)}_{2u}$ 357 cm$^{-1}$ IR mode is decreasing, the E$^{(1)}_{u}$ 276 cm$^{-1}$ IR mode is surprisingly increasing as a function of sample thickness. Both modes show frequency changes and an increase of the width, consistent with that of Raman active phonons.

\subsection{Relative Raman intensity and optical constants}

\begin{figure}
    \includegraphics[width=\linewidth]{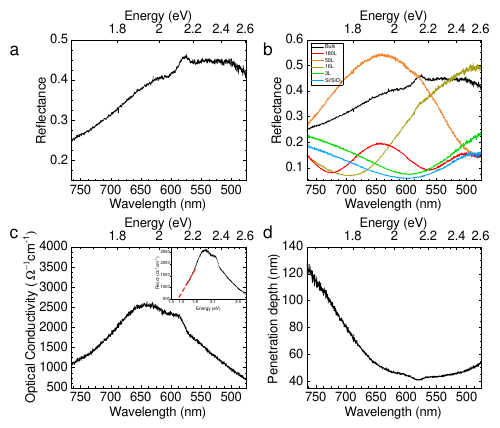}
    \caption{\textbf{Reflectance of Bulk NiGa$_2$S$_4$.} (a) Reflectance spectra of bulk NiGa$_2$S$_4$ (b) Reflectance spectra of bulk plotted in comparison with NiGa$_2$S$_4$ exfoliated flakes of different thickness (c) Calculated real part of optical conductivity of bulk NiGa$_2$S$_4$ (d) Calculated penetration depth of bulk NiGa$_2$S$_4$}
    \label{Fig.reflectance}
\end{figure}

In order to get experimental measure of the energy gap in NiGa$_2$S$_4$ and understand a non-linear dependence of the Raman scattering intensity on the thickness of the flakes, we probed optical constants of NiGa$_2$S$_4$ from 1.3 to 3.5 eV. In general, Raman scattering cross section is proportional to the scattering volume $V_s$ (see, for example, Ref.~\cite{HayesLoudon}):

\begin{equation}
    \frac{\delta^2\sigma}{\delta\Omega\delta\omega_s}=\frac{\omega_I\omega_S^3 V_s v n_S}{16\pi^2c^4 n_I}\bra{} \epsilon_S^i\epsilon_i^j\delta\chi^{ij}|^2\rangle_{\omega}
\end{equation}

where $\omega_I$ and $\omega_S$ are excitation and scattered light frequencies, $n_I$ and $n_S$ are refractive indexes of the excitation and scattered light, $\bra{} \epsilon_S^i\epsilon_i^j\delta\chi^{ij}|^2\rangle_{\omega}$ is fluctuations of susceptibility resulting in Raman scattering. Naturally, if the thickness of flakes decreases below the penetration depth of the laser, a decrease of Raman intensity is expected. We have measured R($\omega$) reflectivity spectra of NiGa$_2$S$_4$ bulk and films to get information about frequency dependent penetration depth and understand the Raman intensity change with thickness.

Reflectivity spectrum of bulk NiGa$_2$S$_4$ in the range of 1.6 to 2.6 eV is shown in Fig.~\ref{Fig.reflectance}(a). A Kramers-Kronig analysis allows us to estimate the optical conductivity as well as the penetration depth of the laser into the bulk of the sample, Fig.~\ref{Fig.reflectance}(c) and (d) respectively. A penetration depth of about 50 nm for 647.1 nm excitation line (Fig.~\ref{Fig.reflectance}(c)) explains the decrease of the intensity of Raman scattering for flakes of thickness lower than 50 layers. Indeed, the flakes of the thickness below 50 layers become transparent, as demonstrated in the Fig.~\ref{Fig.reflectance}(b). For these thin flakes we observe reflectivity determined by the interference of the light in the layered system of Si, 300 nm thick SiO$_2$ substrate, and NiGa$_2$S$_4$ flakes. 

It also points out that the further increase of the Raman response for 9-layer flakes compared to thicker ones (see Fig.~\ref{Fig.Raman}) is determined by the change of the electronic structure, which would renormalize  $\bra{} \epsilon_S^i\epsilon_i^j\delta\chi^{ij}|^2\rangle_{\omega}$

Reflectivity  data also allow us to estimate an optical gap of about 1.5 eV for bulk NiGa$_2$S$_4$ from the linear extrapolation of the edge of optical conductivity to zero~\cite{Zanatta2019}, see the inset in Fig.~\ref{Fig.reflectance}(c).

\subsection{Electronic Structure Calculations for  NiGa$_2$S$_4$}\label{structure}

\begin{figure}
    \includegraphics[width=\linewidth]{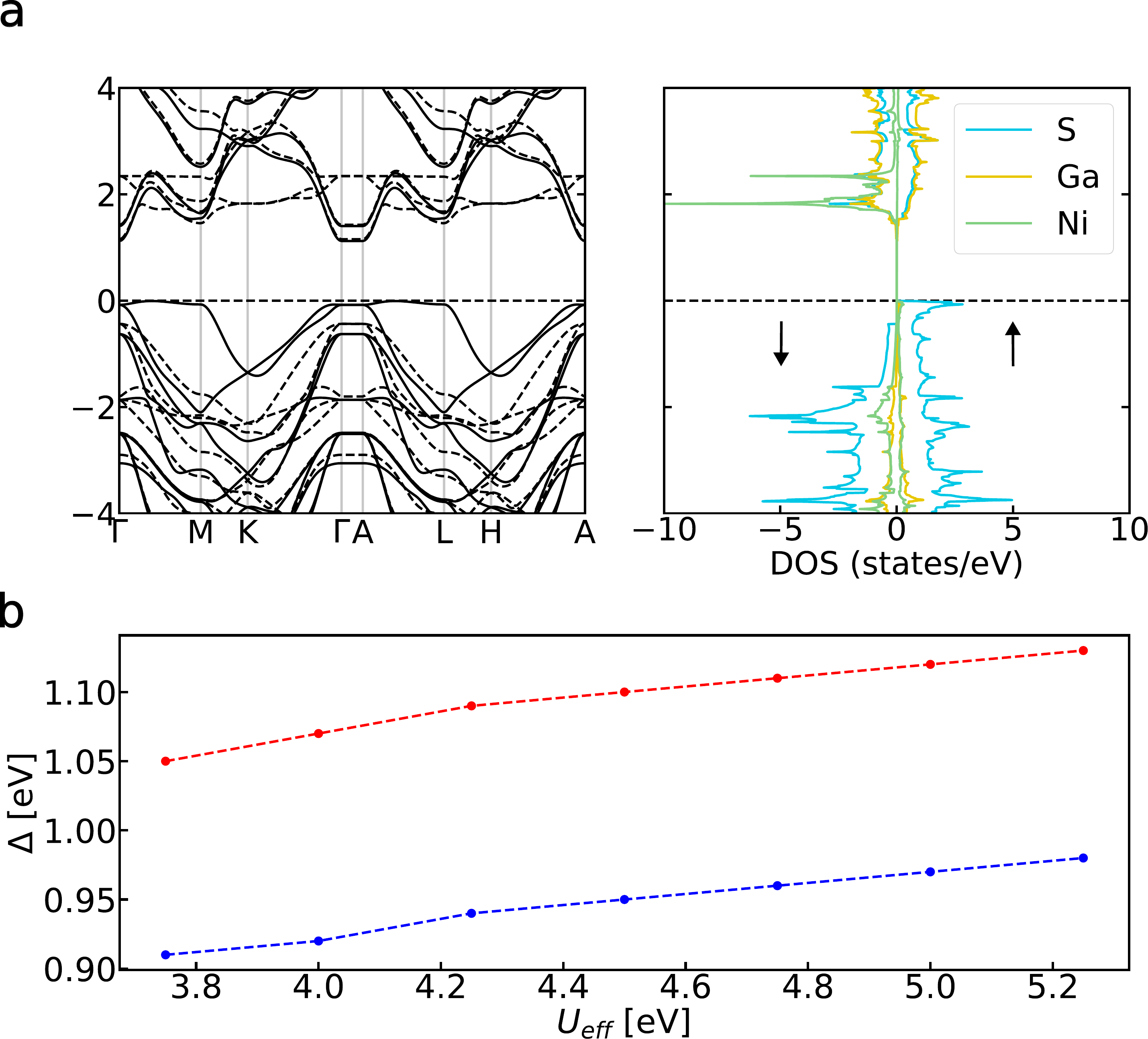}
    \caption{\textbf{Electronic properties and band gap evolution of NiGa$_2$S$_4$ as function of $U_{\rm eff}$ for the bulk and monolayer structure.} (a) Calculated band structure and atom resolved density of states (DOS) for a FM state with $U_{\rm eff}$ = 5.25 eV, with  $U_{\rm eff}= U-J_{\rm H}$. The black solid bands refer to the spin up bands and the dashed to spin down. The arrow up and down indicate the spin up and down channel in the DOS. (b) Blue points show the bulk electronic gap and red points the gap for the monolayer as function of $U_{\rm eff}$ and $J_\text{H}$ = 0.75 eV. The simulations include ferromagnetic ordered magnetic moment on the Ni sites. The dashed lines are a guide for the eye.}
    \label{fig:electronic-properties}
\end{figure}

While the band structure of few-layer and monolayer NiGa$_2$S$_4$ has not been systematically studied, we know from structurally similar materials such as 1T-MoS$_2$ that the band structure can be highly dependent on the number of layers of the material~\cite{MoS2}. Given this, the band structure of the bulk versus monolayer should be discussed.
Therefore, we performed density functional theory (DFT) calculations within FPLO~\cite{fplo} using the GGA+U~\cite{GGA,Koepernik_2009_atomic_limit} exchange correlation functional for bulk and monolayer NiGa$_2$S$_4$ and determined the size of the electronic band gap $\Delta$.
In Fig.~\ref{fig:electronic-properties} we show the  spin-polarized bands and the corresponding atom resolved density of states of monolayer NiGa$_2$S$_4$ for the case of $U_{\rm eff}$ = 5.25 eV, with $U_{\rm eff} = U - J_H$ and $J_H$ = 0.75 eV, as obtained by GGA+U (see Methods section). The valence bands are predominantly of S 3p character hybridized  with  Ni-3d states, while the conduction bands are mainly of Ni-d and Ga 4s character separated by a gap of 1.5 eV.
We observe only small changes of the electronic structure between the bulk~\cite{Serra2023} and the monolayer simulations due to the very weak interaction between layers, confirming the 2D nature of the material. Since the Coulomb correction U as implemented in GGA+U  can be considered as a free parameter, we show in Fig.~\ref{fig:electronic-properties}(b) the dependence of the calculated electronic band gap. We find that the energy gap increases from bulk to the monolayer structure. 
At the level of GGA we obtain a energy gap of 0.12~eV for the bulk, which increases to 0.18~eV for the monolayer.
Including Coulomb corrections on the Ni 3d orbitals at the level of GGA+U with $J_\text{H} = 0.75$ eV and $U$ in a range between 4.5 eV and 6.0 eV
-- which are usual values for Ni d orbitals -- the energy gap increases significantly. 
As shown in Fig.~\ref{fig:electronic-properties}(b) there is a gap difference between the bulk and monolayer of  $\approx$ 0.15 eV for the same $U_{\rm eff}$ value.
However, since the screening effects in the monolayer are smaller, the expected Coulomb correction  U should be larger in the monolayer than in the bulk.
This will lead  to an even larger band gap $\Delta$ difference between the two systems.
This is in agreement with Raman intensities, which increase relative to the bulk  at thicknesses at least below 9 L (see Fig.~\ref{Fig.Raman}), testifying for an increase of electronic band gap and decrease of screening effects from bulk to few-layer systems. We note that the obtention of band gaps from DFT is very sensitive to the choice of basis sets and exchange-correlation functionals used for the calculations.  Previous DFT results on the energy spectra of bulk NiGa$_2$S$_4$ reported band gaps varying around 1.2 - 1.95 eV~\cite{Bandstructure,Serra2023}.  These numbers are compatible with the values obtained in the present study. 

A bandgap of about 1.5 eV was experimentally confirmed using reflectance measurements on the bulk crystal and is estimated from the optical conductivity spectra in Fig.~\ref{Fig.reflectance}(c).


\subsection{Possibility of doping due to proximity of a metal: Calculations}

In order to check the possibility of doping monolayer NiGa$_2$S$_4$ through proximity to a metal, we calculated the work function (WF) 
within DFT by using VASP on a few monolayer structures (see TABLE~\ref{tab:work-functions}).
Our results of the WF for graphene, WeSe$_2$ and WS$_2$ are in very good agreement with previously reported results for these systems~\cite{tuning_Yu_2009,thickness_Kim_2021,strong_britnell_2013,Methane_jang_2019,strain_lanzillo_2015}. We use this agreement  as a benchmark for NiGa$_2$S$_4$ where no previous predictions or experiments are reported, to the best of our knowledge.
The estimated WF for NiGa$_2$S$_4$ is $\approx$ 4.24~eV which is close to the WF value of graphene.
From these results we expect that the charge transfer between NiGa$_2$S$_4$ and graphene will be very small, but should be significant
between NiGa$_2$S$_4$ and WS$_2$.
\begin{table}[h]
\centering
\begin{tabular}{c|c|c}
 & WF-VASP[eV] & WF-Others[eV] \\ \hline \hline
graphene  & 4.20 & 4.30~\cite{graphene_Rutkov_2020} \\
NiGa$_2$S$_4$     & 4.24 & - \\
WSe$_2$ monolayer & 4.38 & $\approx$ 4.30~\cite{thickness_Kim_2021,strong_britnell_2013,Methane_jang_2019} \\
WS$_2$ monolayer  & 4.81 & $\approx$ 4.70~\cite{thickness_Kim_2021,strong_britnell_2013},5.89~\cite{strain_lanzillo_2015}
\end{tabular}
\caption{Work function calculated within DFT by using VASP. For NiGa$_2$S$_4$ we used spin-polarized GGA+U, all other systems are computed at the level of non spin-polarized GGA. Right column displays direct comparison with previously reported results.}\label{tab:work-functions}
\end{table}

\begin{figure}[h!]
    \centering
    \includegraphics[width=\linewidth]{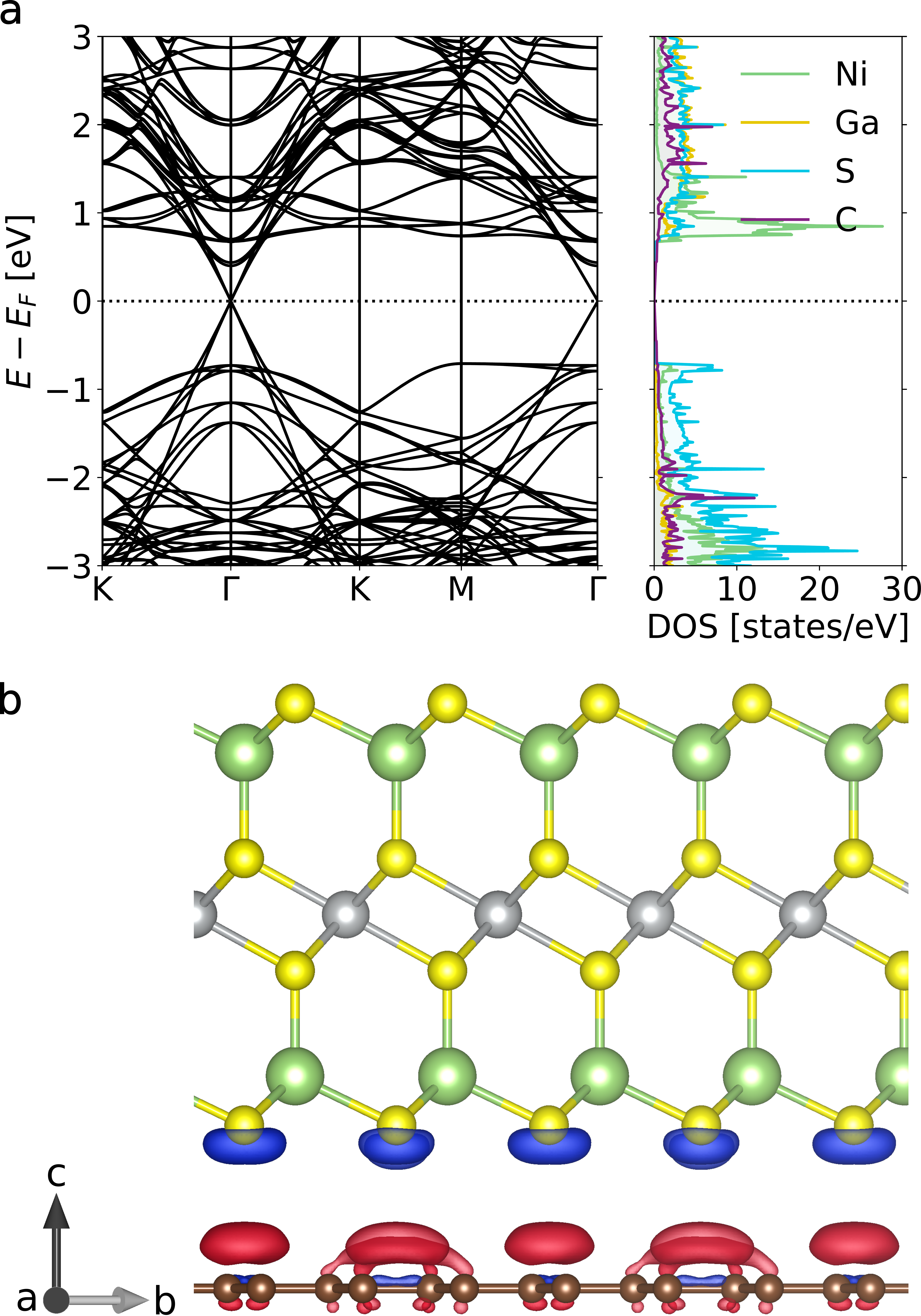}
    \caption{\textbf{Electronic properties for the combined relaxed heterostructure NiGa$_2$S$_4$/graphene}. (a) Non-spin-resolved plot of the calculated electronic bands and density of states (DOS) within GGA+U in VASP, using $U_{\rm eff} = U-J = 4.0$ eV for a ferromagnetic state. The Dirac cone of graphene at the \textbf{$\Gamma$} point is not shifted significantly above the Fermi level indicating that there is very small charge transfer from graphene to the NiGa$_2$S$_4$ layer. (b) The charge difference $\delta$ plot with charge isosurface 1.5$\times$10$^{-4}$ eV/a.u.$^3$. Blue and red charge isosurface refer to charge accumulation and deficit, respectively. }
    \label{fig:electronic-graphene}
\end{figure}

In a next step, we also calculated the electronic properties of NiGa$_2$S$_4$ on graphene following a similar procedure as in previous
bilayer relaxations~\cite{biswas2019,balgley2022,crippa2024} and including in our DFT calculations van der Waals corrections (see Methods section). The relaxed monolayer NiGa$_2$S$_4$ structure  undergoes a tensile strain (extended) by 2\% in the a-b plane compared to the bulk system, due to the lattice mismatch with graphene.
The distance between monolayer NiGa$_2$S$_4$ and graphene in the relaxed structure is 3.414~\AA.
The triangular lattice built by the Ni sites is preserved with a Ni-Ni distance $d_{\rm Ni-Ni} = 3.689$~\AA~which is close to the bulk distance of 3.627~\AA.
Our Bader analysis shows very small charge transfer in the combined system  with $\delta \rho_{\rm NiGa_2S_4}  = -0.0007$ e/f.u. and carbon of $\delta \rho_{\rm C}  = 0.00002$ e/C (see Fig.~\ref{fig:electronic-graphene}(b)).
This can be also observed in Fig.~\ref{fig:electronic-graphene}(a), where the Dirac cone from the graphene layer is not shifted away from the Fermi level.  This confirms that there is no significant charge transfer in the NiGa$_2$S$_4$/graphene bilayer, as suggested by the WF analysis.
Note that due to the considered 3$\times$3 supercell for graphene, the \textbf{K} point of graphene is folded at the \boldsymbol{$\Gamma$} point~\cite{zhou2013_3n}.

\section{Discussion and Conclusions}\label{sec3}

Our results demonstrate that triangular lattice antiferromagnet NiGa$_2$S$_4$ can be exfoliated down to at least 2-layer flakes, and that flakes of different thickness can be successfully characterized by Raman spectroscopy. As expected, sulfur Raman active phonons broaden and shift to lower frequencies when reaching the thinnest number of layers. While phonons related to the movement of ``outer'' sulfurs of the Ga environment were expected to show strong changes with the exfoliation, our results show that phonons related to the movement of ``inner'' sulfurs of the Ni octahedral environment actually showed stronger changes with exfoliation. The comparison of AFM and Raman characterization of thin flakes presented in this work will allow to further use the Raman scattering characterization for thin flakes of NiGa$_2$S$_4$.

An unexpected effect of the exfoliation of NiGa$_2$S$_4$ to relatively thick layers around 180L (277 nm) is the increase in Raman intensity and shift of all the phonon frequencies from the bulk. A similar effect is seen in 50L (77 nm) where all the phonon frequencies shift to higher values and decrease in their width, with the largest effect on the phonon at 450 \cm. The effect is similar to what would be expected on a decrease of structural disorder. The natural explanation would be a decrease of the number of stacking faults in the flakes~\cite{Rohmfeld1998}. As characterized by Raman spectra, flakes of the thickness of 50L and less are free of stacking faults. This also suggests a thickness of a maximum single domain in NiGa$_2$S$_4$ along the $c$ axis of about 50L.

While the effect of stacking faults decreases upon exfoliation, the intensity of IR-active phonons in Raman spectra is persistent down to at least 3L-thick flakes. We previously suggested that the intensity of IR bands appears due to a loss of local inversion center as a result of sulfur vacancies~\cite{Valentine2020}, which is consistent with their presence in exfoliated flakes as well. 

The band structure calculations confirm that NiGa$_2$S$_4$ is a Mott insulator. The gap estimated to be about 1.5 eV from reflectivity measurements of the bulk sample is confirmed by the calculations, within the limitations of DFT. The experimentally observed increase of Raman scattering intensity in the thin layers is understood as an increase of the electronic band gap in a single layer influenced by a decrease in screening effects, as reproduced by the electronic structure calculations. 

Finally, we explored the possibility of doping  NiGa$_2$S$_4$ through charge transfer in proximity to a metal layer, as a part of a vdW heterostructure assembly. Our calculations show that almost no charge transfer is expected in proximity to  graphene but may be significant between NiGa$_2$S$_4$ and WS$_2$.

In conclusion, we introduced a new magnetic vdW material NiGa$_2$S$_4$, which is, in the bulk, a S=1 frustrated antiferromagnet on a triangular lattice. We demonstrated the possibility of mechanical exfoliation and characterization of thin flakes with AFM and Raman scattering spectroscopy. Thinner flakes of this material are shown to be free of stacking faults, with the single-stacking domain thickness estimated to be of the order of 50L. We complemented our study with electronic structure calculations for bulk and monolayer NiGa$_2$S$_4$= as well as heterostructures of NiGa$_2$S$_4$ on graphene. We find that while doping such a material is a potentially interesting way to achieve exotic conducting properties, doping of NiGa$_2$S$_4$ by proximity to standard vdW metallic layers may not be the best route to achieve conductivity in a single layer of this magnetic material, except for the case of WS$_2$.

\section{Methods}\label{sec4}

\subsection{Sample Preparation}\label{prep}

NiGa$_2$S$_4$ single crystals were grown using the method outlined in Ref.~\cite{Nambu2008synthesis}. The resulting crystals are thin plates with the most developed surface parallel to $ab$ plane measuring up to 3 mm by 3 mm and thickness 10~$\mu$m.

The traditional mechanical exfoliation method was implemented to exfoliate flakes of NiGa$_2$S$_4$ from the bulk single crystal. This method applies mechanical force to the bulk of the material sandwiched between a piece of blue nitto tape and peels layers away by repeatedly pulling the tape apart. After this exfoliation, the tape is pressed down onto a chemically cleaned Si/SiO$_2$ wafer and peeled away. Wafers were cleaned by sonicating in heated acetone for 15 minutes followed by sonicating in IPA for 10 minutes and then drying with pressed air. 

\subsection{Atomic Force Microscopy characterization}\label{AFM method}

AFM characterization was carried out using an Asylum research AFM at room temperature under ambient conditions. AFM data was analyzed with Gwyddion software.  

\subsection{Raman spectroscopy characterization}\label{Raman method}

Raman spectra was measured on a micro-Raman set-up using a T64000 Horiba Jobin-Yvon spectrometer equipped with an Olympus microscope. Raman scattering was excited with the 514.5 nm and 647.1 nm excitation wavelength lines of an Ar$^+$-Kr$^+$ gas laser. A laser probe of 2 $\mu$m in diameter was used. For the spectra of NiGa$_2$S$_4$ flakes with thickness below the penetration depth, the resulting spectra are the superposition of NiGa$_2$S$_4$ flake and Si wafer. In this case, in order to separate spectra of NiGa$_2$S$_4$ flakes, Si background spectra were measured on the same wafer as exfoliated flakes and subtracted from  the spectra of the samples. The resulting spectra of the flakes are presented in the manuscript.  

\subsection{Reflectance measurements and calculations of optical constants}

Reflectance spectra at the normal angle of incidence was measured using the Olympus microscope and T64000 Horiba Jobin-Yvon spectrometer in a broad band mode, with a broadband white light source used as a source of radiation. To obtain absolute values of reflectivity, reflectance of the sample was compared to the reflectance of a silver mirror with the know reflectivity values. To obtain optical constants from reflectivity R($\omega$) spectra we performed Kramers-Kronig transformation. Low frequency extrapolation by a constant was used, based on the fact that NiGa$_2$S$_4$ is a non-conducting material. Typical high frequency extrapolation as R($\omega$) = $\frac{A}{\omega^2}$ was used.  

\subsection{Bulk and monolayer electronic properties calculations}

The electronic properties of the bulk structure are calculated in the space group 164 with $a=b=$ $3.6268$~\AA~and $c$ = $12.0018$~\AA~ ($\alpha$=$\beta$=$90^\circ$, $\gamma$=$120^\circ$)~\cite{stackingfaults}.
The monolayer calculations are performed on the same crystal structure, where a void of 27~\AA~between the NiGa$_2$S$_4$ layers is introduced. 
All results are obtained within density functional theory (DFT) \textit{via} the full potential local orbital (FPLO)~\cite{fplo} code version 21.00-61.
As exchange functional we select the generalized gradient approximation~\cite{GGA} (GGA) and include effective Coulomb correction on the strongly corrected 3d Ni orbitals \textit{via} the DFT+U scheme using the atomic limit method~\cite{Koepernik_2009_atomic_limit}, with $J_\text{H} = 0.75$ eV and $U=4.5$ eV.
All calculation are spin-polarized with ferromagnetic ordered moments on the Ni sites.
The bulk calculations are performed on a 12$\times$12$\times$4 k-grid (12$\times$12$\times$1 for the monolayer) and density convergence criterion of $10^{-6}$.

\subsection{NiGa$_2$S$_4$/Graphene heterostructure calculations}

For the bilayer calculations, as a starting point we choose NiGa$_2$S$_4$ crystal structure in the space group 164 with $a=b=$ $3.6268$~\AA~and $c$ = $12.0018$~\AA~ ($\alpha$=$\beta$=$90^\circ$, $\gamma$=$120^\circ$)~\cite{stackingfaults},
and extract a monolayer from the bulk system which is placed on a 3$\times$3-graphene layer with C-C distances $d_{\rm C-C}$ = 1.420~\AA. 
This initial geometry is relaxed within DFT using the VASP simulation package~\cite{Kresse_VASP} version 6.3.0, where the graphene layer is fixed but atoms in the NiGa$_2$S$_4$ monolayer are free to move in any direction.
As exchange correlation functional we choose the generalized gradient approximation~\cite{GGA} (GGA), include Coulomb corrections on the Ni 3d orbitals \textit{via} the Dudarev~\cite{Dudarev_LDA+U} DFT+U scheme with $U_{\rm eff} = 4.0$~eV and Van der Waals corrections \textit{via} the DFT+D2 method of Grimme~\cite{grimme_vdw}.
The structural simulations have been performed with a plane wave cut-off of 600 eV for the expansion of the basis set, on a 9$\times$9$\times$1 k-mesh including a ferromagnetic moments on the Ni sites, until the forces for each atom in all directions decrease down to $10^{-3}$~eV/\AA~with energy convergence criterion of $10^{-7}$~eV.
The electronic properties have been calculated using higher dense 12$\times$12$\times$1 k-grid.
To calculate the charge transfer we use Bader analysis as implemented in Reference~\cite{HENKELMAN_a_fast_2006}.

Raw data used and/or analyzed in this study is available from the corresponding author on request.

\section{Acknowledgements}

Work at JHU was supported as part of the Institute for Quantum Matter, an Energy Frontier Research Center funded by the U.S. Department of Energy, Office of Science, Basic Energy Sciences under Award No.~DE-SC0019331. JV is grateful to NSF AGEP-GRS program, a supplement to a current DMR grant \# 2004074 for the support. AR amd RV acknowledge support by the Deutsche Forschungsgemeinschaft (DFG, German Research Foundation)  through 
TRR 288 -- 422213477 (project A05, B05). T.J.K. acknowledges funding from a National Science Foundation (DMR-1848046) CAREER grant. T.J.K. also acknowledges funding for this study by the Young Faculty Award program of the Defense Advanced Research Projects Agency (DARPA) and by the Army Research Office under the grant W911NF-21-1-0351. RDG acknowledges funding from the ARCS Foundation Metro Washington Chapter as a JCM Foundation Scholar. T.H. and S.N. acknowledge support of JST-MIRAI Program (No. PMJMI20A1) and JST-ASPIRE(JPMJAP2317) .

\bibliography{references}

\end{document}